\documentclass[12pt]{article}
\usepackage{graphicx}

\usepackage{amsmath}
\usepackage{amssymb}
\usepackage{amsthm}
\usepackage{latexsym}

\newcommand{\comment}[1]{}

\begin{document}

\date{}

\title{{\LARGE\bf A Note on Edwards' Hypothesis for \\ Zero-Temperature Ising Dynamics}}

\author{
{\bf Federico Camia}
\thanks{Research supported by a Marie Curie Intra-European Fellowship
under contract MEIF-CT-2003-500740.}\,
\thanks{E-mail: camia@eurandom.tue.nl}\\
{\small \sl EURANDOM, P.O. Box 513, 5600 MB Eindhoven, The Netherlands}\\
}

\maketitle

\begin{abstract}
We give a simple criterion for checking the so called Edwards'
hypothesis in certain zero-temperature, ferromagnetic spin-flip
dynamics and use it to invalidate the hypothesis in various examples
in dimension one and higher.
\end{abstract}


\section{Introduction}

In many physical systems, the dynamics at low temperature
or high density is so slow that the system is out of
equilibrium at all practical time scales, so that ordinary
thermodynamics does not apply to those situations.
As a consequence, because of their practical as well as
theoretical interest, non-standard thermodynamics have been
proposed to describe out-of-equilibrium systems.
(The literature concerning such systems is huge --
for a few examples across the spectrum, see for
instance~\cite{bray,gc,ckp,bckm,fin,ruelle,mk,cdsn,babg2},
and references therein.)

In the context of the slow compaction dynamics of granular
materials, Edwards~\cite{ed1,ed2,me} proposed to compute
thermodynamic quantities by means of a flat ensemble average
over all the blocked configurations of grains with prescribed
density, leading to a natural definition of configurational
temperature.
Since the approach is not justified from first principles,
its validity has to be tested with specific models and
experiments (see, for example, \cite{mk} and references therein).

There are other situations in which the idea of an effective
temperature is very appealing, and it is tempting to extend
the range of applicability of Edwards' idea to other
systems with a large number of ``metastable'' states.
In this broader context, what came to be known as Edwards'
hypothesis consists in assuming that all the metastable states
in which a system can be trapped are equivalent for the dynamics.
This corresponds, as before, to computing thermodynamic quantities
using a flat ensemble average over all the metastable states.

A situation where this prescription makes sense is represented
by glassy dynamics, which are often described as a slow motion in
a complex energy (or free energy) landscape, with many ``valleys''
separated by barriers.
Several approaches have been proposed to make this heuristic picture
more precise, and different notions of ``valleys'' have appeared,
but no general and unambiguous definition of metastable states has
appeared yet.
Nonetheless, once these valleys are appropriately defined,
one can estimate their number ${\cal N}_E(N)$ at fixed energy
(or free energy) density $E$ for a fixed size $N$ of the system.
${\cal N}_E(N)$ generally grows exponentially with the size of the system:
\begin{equation}
{\cal N}_E(N) \sim \exp{(N \, S_c(E))},
\end{equation}
where $S_c(E)$ is a \emph{configurational entropy} or \emph{complexity}.
The key question concerning the dynamics is whether all valleys
play the same role, or whether they have different \emph{dynamical}
weights, according to their basin of attraction.
This question arises, for instance, when a system is instantaneously
quenched into the glassy (low temperature) phase, starting from a
disordered (high temperature) configuration.

In this context, Edwards' hypothesis is valid for some mean-field
models, where valleys are know to be explored with a flat
measure~\cite{fv}.
One can therefore define a thermodynamics based on the flat
ensemble over valleys and a \emph{configurational temperature}
$T_c$ by
\begin{equation} \label{c-temp}
\frac{1}{T_c} = \frac{\text{d}S_c}{\text{d}E}.
\end{equation}

Besides the mean-field case, another situation where valleys
can be unambiguously defined is the zero-temperature limit,
where no energy barrier can be crossed and the valleys correspond
to the blocked configurations under the chosen dynamics.
Models with constrained dynamics have been extensively studied
in one dimension (see~\cite{dsgl} and references therein), where
Edwards' hypothesis has been tested using exact results on the
statistics of blocked configurations, as well as numerical simulations.
The results obtained so far seem to imply that the hypothesis cannot
be applied to zero-temperature spin-flip dynamics~\cite{dsgl}.

In this paper, we introduce a general criterion for checking
Edwards' hypothesis for \emph{attractive} (this refers to the
ferromagnetic nature of the interaction between spins -- see
Section~\ref{fkg+harris}) symmetric Ising dynamics with initial
configuration chosen from a symmetric distribution that satisfies
the FKG inequality (see~\ref{fkg+harris} and~\cite{harris1,fkg,grimmett}).
As a first application (Section~\ref{check}), we rigorously invalidate
Edwards' hypothesis for the one-dimensional constrained Glauber
dynamics analyzed in~\cite{dsgl} (see also~\cite{dl,ld,pb}), which
we use as a prototype.
Later (Section~\ref{high}), we show how the criterion can be easily
applied to other dynamics in higher dimensions, where the analytic
techniques of~\cite{dsgl} cannot be used and exact results are not
available.


\subsection{FKG Inequality and Harris' Theorem} \label{fkg+harris}

We give here the tools needed to check Edwards' hypothesis for
attractive (see the next paragraph), zero-temperature Ising dynamics.
In the context of an Ising spin model on a lattice $\mathbb L$,
we will call \emph{increasing} an event $\cal E$ such that its
indicator function $I_{\cal E}(\sigma)$ is increasing in the
number of plus spins present in the configuration
$\sigma = \{ \sigma_x \}_{x \in {\mathbb L}}, \, \sigma_x = \pm 1$.
If ${\cal E}_1$ and ${\cal E}_2$ are two increasing events, the
FKG inequality (see, for instance, \cite{fkg,grimmett}) states
that, roughly speaking, the occurrence of ${\cal E}_2$ makes
${\cal E}_1$ more likely, or more precisely, the conditional
probability of ${\cal E}_1$ given ${\cal E}_2$ is larger than
or equal to the probability of ${\cal E}_1$:
\begin{equation} \label{fkg-in}
P({\cal E}_1 \, | \, {\cal E}_2) \geq P({\cal E}_1).
\end{equation}
Many interesting distributions satisfy the FKG
inequality~(\ref{fkg-in}), among them are Gibbs measures
(with some restrictions) and product measures, and in particular
the symmetric Bernoulli product measure from which the initial
configuration of the constrained Glauber dynamics of
Section~\ref{cgd} is chosen (corresponding to a spin
system prepared at ``infinite" temperature).

We will say that a spin-flip dynamics is \emph{attractive}
if, for all vertices $x$ of $\mathbb L$, the rate for the
spin flip $\sigma_x = -1 \to \sigma_x = +1$ is non-decreasing
in the number of plus spins in $\sigma$ (we will consider
only symmetric dynamics, so the roles of plus and minus spins
can be interchanged).
Stochastic Ising models with a ferromagnetic interaction are
examples of attractive dynamics.
A theorem of Harris~\cite{harris2,liggett} states that
attractive dynamics preserve the FKG property, i.e.,
if one starts with a measure $P_0$ that satisfies
the FKG inequality and applies to the spin system an
attractive dynamics, the measure $P_t$ describing the
spin system at time $t$ still satisfies the FKG inequality.

In particular, this result can be applied to the
constrained Glauber dynamics of Section~\ref{cgd}
to deduce that the limiting (as $t \to \infty$)
measure $P_{\infty}$ satisfies the FKG inequality.

\section{A Constrained Glauber Dynamics in 1D} \label{cgd}

We consider the following ferromagnetic Ising chain with single
spin-flip (Glauber) dynamics (see~\cite{dsgl} for more details)
with (formal) Hamiltonian
\begin{equation} \label{hamiltonian}
{\cal H}(\sigma) = - \sum_{n \in {\mathbb Z}} \sigma_n \sigma_{n+1}
\end{equation}
and flipping rates $W$ determined by the energy difference between
the configurations after and before the proposed move, i.e.,
\begin{equation}
W(\sigma_n \to - \sigma_n) = {\cal W}_{\delta {\cal H}},
\end{equation}
with
\begin{equation}
\delta {\cal H} = 2(\sigma_{n-1} + \sigma_{n+1}) \sigma_n \in \{-4,0,4 \}.
\end{equation}

Detailed balance with respect to~(\ref{hamiltonian}) at inverse temperature
$\beta$ imposes the condition:
\begin{equation}
\frac{{\cal W}_4}{{\cal W}_{-4}} = e^{- 4 \beta}.
\end{equation}
Choosing time units so that ${\cal W}_{-4} = 1$, we have
${\cal W}_4 = e^{- 4 \beta}$.
As in~\cite{dsgl}, we are interested in the zero-temperature
case, so that ${\cal W}_4 = 0$.
The rate ${\cal W}_0$ remains a free parameter, which we choose
to be zero (corresponding to the zero-temperature limit of the
Glauber dynamics).
Setting ${\cal W}_0 = 0$ corresponds to allowing only spin flips
that lower the energy, therefore the only possible moves, happening
with rate $1$, are flips of plus spins surrounded by minus spins or
minus spins surrounded by plus spins:
\begin{equation}
-+- \to ---, \, \, \, \, \, \, \, +-+ \to +++.
\end{equation}
The blocked configurations (i.e., the absorbing states of the dynamics)
are those where the unsatisfied bonds (i.e., bonds between spins
of opposite sign) are isolated.

We consider the \emph{deep-quench} situation, where the system
is prepared at infinite temperature and the temperature is then
decreased to zero instantaneously.
This corresponds to an initial configuration chosen randomly from
a symmetric Bernoulli product measure, i.e., with
\begin{equation} \label{initial}
\left\{ \begin{array}{ll}
        \sigma_n(0) = +1 & \mbox{with probability } 1/2 \\
        \sigma_n(0) = -1 & \mbox{with probability } 1/2
                             \end{array}
                     \right.
\end{equation}
where $\sigma_n(t)$ is the value of the spin $\sigma_n$ at time $t$.
We call $P_t$ the distribution of the spin configuration
$\sigma(t) = \{ \sigma_n(t) \}_{n \in {\mathbb Z}}$ at time $t$, and
denote by $P_{\infty}$ the limiting distribution
obtained as $t \to \infty$.

\subsection{Checking Edwards' Hypothesis} \label{check}

In~\cite{dsgl}, $P_{\infty}$ is compared to the uniform
distribution $P_{unif}$ on blocked configurations,
corresponding to an ensemble where \emph{all} blocked
configurations have the \emph{same} weight.
The conclusions reached there, using exact results on the
statistics of the blocked configurations reached by the system,
invalidate Edwards' hypothesis for this particular model.

Here, we confirm those results by rigorously proving that
$P_{\infty} \neq P_{unif}$, but the main goal of this
section is to introduce, via a simple specific example,
a general criterion for comparing the limiting distribution
$P_{\infty}$ of a spin system subjected to an attractive
dynamics to the uniform distribution $P_{unif}$ on the absorbing
states. 
The general strategy is described in Section~\ref{gen-strat},
where we also give further applications.

Consider all blocked configurations of the spin chain such
that $\sigma_{\pm 2} = \sigma_{\pm 3} = +1$.
It is easy to see that such blocked configurations are of
only four different types:
\begin{enumerate}
\item[A] \hskip.5cm $\sigma_{-1} = \sigma_0 = \sigma_1 = +1$ \hskip1.9cm
\hskip1cm $\ldots +++++++ \ldots$
\item[B] \hskip.5cm $\sigma_{-1} = +1, \,\,\, \sigma_0 = \sigma_1 = -1$ \hskip1cm
\hskip1cm $\ldots +++--++ \ldots$
\item[C] \hskip.5cm $\sigma_{-1} = \sigma_0 = -1, \,\,\, \sigma_1 = +1$ \hskip1cm
\hskip1cm $\ldots ++--+++ \ldots$
\item[D] \hskip.5cm $\sigma_{-1} = \sigma_0 = \sigma_1 = -1$ \hskip1.9cm
\hskip1cm $\ldots ++---++ \ldots$
\end{enumerate}
Under the uniform distribution on blocked configuration,
the occurrence of each type has equal probability; therefore,
conditioned on having $\sigma_{\pm 2} = \sigma_{\pm 3} = +1$,
\begin{equation}
P_{unif}(\sigma_0 = +1 \, | \, \sigma_{\pm 2} = \sigma_{\pm 3} = +1)
= P_{unif}(A \, | \, \sigma_{\pm 2} = \sigma_{\pm 3} = +1) = 1/4.
\end{equation}

On the other hand, Harris' Theorem (see Section~\ref{fkg+harris})
applied to this specific dynamics implies that $P_{\infty}$
satisfies the FKG inequality, so that we have
\begin{equation} \label{fkg}
P_{\infty}(\sigma_0 = +1 \, | \, \sigma_{\pm 2} = \sigma_{\pm 3} = +1)
\geq P_{\infty}(\sigma_0 = +1) = 1/2,
\end{equation}
where the equality follows from the $\pm$ symmetry of the dynamics
and the initial distribution.
The last two equations show that $P_{\infty}$ cannot be the uniform
distribution $P_{unif}$.

\section{The General Strategy} \label{gen-strat}

The strategy we used for the constrained Glauber dynamics of the
previous section can be generalized to any attractive, symmetric
Ising dynamics with locally stable configurations (later, we will
also give an application where there are no locally stable configurations
-- see Example~4 in Section~\ref{high}), with initial configuration
chosen from a symmetric distribution that satisfies the FKG inequality.
For simplicity, we restrict our attention to nearest neighbor models;
in this context, by the existence of locally stable configurations
we mean that there are \emph{finite} subsets $G$ of the lattice $\mathbb L$
such that, if $\sigma_x(t_0) = +1 \,\, (-1) \,\,\, \forall x \in G$,
then $\sigma_x(t) = +1 \,\, (-1) \,\,\, \forall x \in G, \,\,\,
\forall t>t_0$.
When this is the case, we say that the spins in $G$ are
\emph{stable} and we call $G$ a \emph{stable set}.
If $G$ is a smallest set with this property (there could
be more than one, with different shapes), we call it a
\emph{minimal stable set}.

Some more notation is needed before we can proceed with the general
strategy and further applications.
Given a subset $\Lambda$ of $\mathbb L$, we call \emph{exterior boundary}
$\partial_e \Lambda$ of $\Lambda$ the set of vertices $x \notin \Lambda$
that are adjacent to a vertex in $\Lambda$, and \emph{interior boundary}
$\partial_i \Lambda$ of $\Lambda$ the set of vertices $x \in \Lambda$
that are adjacent to a vertex not in $\Lambda$.

We are now ready to explain the general strategy;
in the next section we will illustrate it with some examples.
Let $G_1$ and $G_2$ be two distinct minimal stable sets both
containing the origin ($0 \in G_1 \cap G_2$) and denote by
$G = G_1 \cup G_2$ their union.
Let $L$ be a (finite) stable set such that $G \cap L = \emptyset$
and $\partial_e G \subset L$ (in words, $G$ is ``surrounded" by $L$).
$G_1$, $G_2$ and $L$ should be chosen so that
$\{ G \setminus G_1 \} \cup L$ and $\{ G \setminus G_2 \} \cup L$
are stable sets.
Notice that, since $G_1$ and $G_2$ are minimal stable sets,
$G \setminus G_1$ and $G \setminus G_2$ are smaller than any
minimal stable set and therefore are not stable sets.

Now it is easy to convince oneself that, conditioned on the spins
in $L$ all being plus, there are only four possible types of blocked
configurations:
\begin{enumerate}
\item All the spins in $G$ are plus.
\item All the spins in $G$ are minus.
\item The spins in $G_1$ are minus and those in $G \setminus G_1$
are plus.
\item The spins in $G_2$ are minus and those in $G \setminus G_2$
are plus.
\end{enumerate}
This implies that, conditioned on all the spins in $L$ being plus,
the uniform distribution on stable configurations assigns probability
$1/4$ to the event that the spin at the origin is plus (corresponding
to case 1 above).

On the other hand, if we consider a symmetric, attractive dynamics
with initial configuration chosen from a symmetric Bernoulli
product measure, conditioned on the same event (all the spins in
$L$ being plus), the limiting distribution $P_{\infty}$ must assign
probability at least $1/2$ to the event that the spin at the origin
is plus, which clearly shows that $P_{\infty}$ cannot be the uniform
distribution.

\subsection{Higher Dimensional Examples} \label{high}

Here we present some examples in dimension higher than one
where we can use the method described above to rule out the
uniform distribution.
All we have to do is choose the sets $G_1$, $G_2$, and $L$
appropriately.
We will consider zero-temperature dynamics such that a spin
flips at rate $1$ if it disagrees with a strict majority of
its neighbors and at rate $0$ otherwise.
As in the example of Section~\ref{cgd}, we will always start
with a Bernoulli symmetric product measure (see~(\ref{initial})),
corresponding to the deep-quench situation.
We note that exact results are usually not available for higher-dimensional
models, which limits the range of applicability of the methods used
in~\cite{dsgl} to one-dimensional models.


\bigskip

\begin{figure}[!ht]
\begin{center}
\includegraphics[width=6cm]{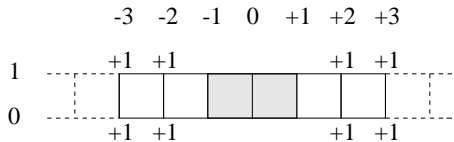}
\caption{The increasing event $\cal E$ used in Example~1; $\sigma_n$
are the spins in the lower row ${\mathbb Z} \times \{ 0 \}$ and $\sigma'_n$
those in the upper one ${\mathbb Z} \times \{ 1 \}$.}
\label{fig1}
\end{center}
\end{figure}

\noindent {\bf Example~1}: Zero-temperature dynamics on the ladder ${\mathbb Z} \times \{ 0,1 \}$. \\
\noindent The blocked configurations are such that each spin has at least two
neighbors of the same sign; squares are minimal stable sets.
We choose $G_1$ and $G_2$ to be the sets of vertices of the two squares containing
the origin $\{ 0 \} \times \{ 0 \}$ (the shaded squares in Figure~\ref{fig1}) and
$L$ to be the set of vertices $\{ \pm 2, \pm 3 \} \times \{ 0,1 \}$.

Conditioning on the increasing event
${\cal E} = \{ \sigma_{\pm 2} = \sigma_{\pm 3} = \sigma'_{\pm 2} = \sigma'_{\pm 3} = +1 \}$
(see Figure~\ref{fig1}), it is easy to see that $P_{unif}(\sigma_0=+1 \, | \, {\cal E}) = 1/4$,
because the fact that $\sigma_0=+1$ implies that $\sigma_{\pm 1} = \sigma'_{\pm 1} = +1$,
while there are three possible local blocked configurations with $\sigma_0=-1$.
On the other hand, $P_{unif}(\sigma_0=+1) = 1/2$ by symmetry,
so that $P_{unif}$ does not satisfy the FKG inequality.

\begin{figure}[!ht]
\begin{center}
\includegraphics[width=4cm]{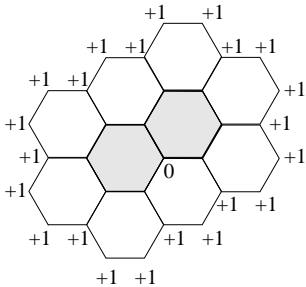}
\caption{The increasing event $\cal E$ used in Example~2.
As indicated, the vertices of the interior boundary
$\partial_i \Lambda$ of $\Lambda$ all have spin $+1$.}
\label{fig2}
\end{center}
\end{figure}

\noindent {\bf Example~2}: Zero-temperature dynamics on the hexagonal lattice. \\
\noindent The blocked configurations are again such that each spin has
at least two neighbors of the same sign; hexagons are minimal stable sets.
Let $\Lambda = G_1 \cup G_2 \cup L = G \cup L$ be the set of vertices of
the portion of hexagonal lattice shown in Figure~\ref{fig2}, where $G_1$
and $G_2$ are the sets of vertices of the two shaded hexagons containing
the origin and $\partial_e G \subset L = \Lambda \setminus G = \partial_i \Lambda$.

Conditioning on the increasing event
${\cal E} = \{ \sigma_y=+1, \forall y \in L = \partial_i \Lambda \}$
that all the spins in $L = \partial_i \Lambda$ are $+1$, it is easy to see
that $P_{unif}(\sigma_0 = +1 \, | \, {\cal E}) = 1/4$, because the fact that
$\sigma_0=+1$ implies that $\sigma_y=+1$ for all $y \in G$, while there are
three possible local blocked configurations with $\sigma_0=-1$.
On the other hand, $P_{unif}(\sigma_0=+1) = 1/2$ by symmetry,
so that $P_{unif}$ does not satisfy the FKG inequality.

\begin{figure}[!ht]
\begin{center}
\includegraphics[width=4cm]{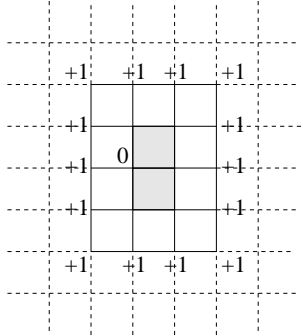}
\caption{The increasing event $\cal E$ used in Example~3.
As indicated, the vertices of the interior boundary
$\partial_i \Lambda$ of $\Lambda$ all have spin $+1$.}
\label{fig3}
\end{center}
\end{figure}

\noindent {\bf Example~3}: Zero-temperature dynamics on ${\mathbb Z}^d$. \\
\noindent For simplicity, we consider the two-dimensional case $d=2$,
but the same reasoning works for all $d \geq 2$.
In two dimensions the blocked configurations are again such that each spin
has at least two neighbors of the same sign; squares are minimal stable sets.
Let $\Lambda = G_1 \cup G_2 \cup L = G \cup L$ be the set of vertices of
the portion of square lattice shown in Figure~\ref{fig3}, where $G_1$
and $G_2$ are the sets of vertices of the two shaded squares containing the
origin and $\partial_e G \subset L = \Lambda \setminus G = \partial_i \Lambda$.

Conditioning on the increasing event
${\cal E} = \{ \sigma_y=+1, \forall y \in L = \partial_i \Lambda \}$
that all the spins in $L = \partial_i \Lambda$ are $+1$, it is easy to see
that $P_{unif}(\sigma_0 = +1 \, | \, {\cal E}) = 1/4$, because the fact that
$\sigma_0=+1$ implies that $\sigma_y=+1$ for all $y \in G$, while there are
three possible local blocked configurations with $\sigma_0=-1$.
On the other hand, $P_{unif}(\sigma_0=+1) = 1/2$ by symmetry,
so that $P_{unif}$ does not satisfy the FKG inequality.

\bigskip

\noindent {\bf Example~4}: Zero-temperature dynamics on the Cayley tree of degree $3$. \\
\noindent This last example is interesting because, contrary to all the previous ones,
there are no locally stable configurations (the only stable structures are doubly-infinite
plus or minus paths).
Nonetheless, the criterion described in this paper can still be used.

\begin{figure}[!ht]
\begin{center}
\includegraphics[width=4cm]{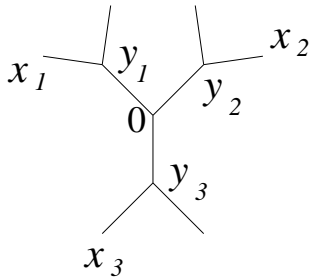}
\caption{A portion of the Cayley tree of degree three.}
\label{fig4}
\end{center}
\end{figure}

With reference to Figure~\ref{fig4}, conditioning on the increasing event
${\cal E} = \{ \sigma_{x_1}=\sigma_{x_2}=\sigma_{x_3}=+1 \text{ and $x_1,x_2,x_3$
belong to doubly-infinite $+1$ paths that do not contain $y_1,y_2,y_3$} \}$,
it is easy to see that, while $P_{unif}(\sigma_0=+1)=1/2$ by symmetry,
$P_{unif}(\sigma_0=+1 \, | \, {\cal E})=1/4$, because if $\sigma_0=+1$, then
$\sigma_{y_1}$, $\sigma_{y_2}$ and $\sigma_{y_3}$ are all forced to be $+1$.
Therefore, once again $P_{unif}$ does not satisfy the FKG inequality.

\section{More on the Constrained Glauber Dynamics}

In~\cite{dsgl}, $P_{\infty}$ is compared to two distributions:
the uniform distribution $P_{unif}$ on all blocked configurations,
already discussed in Section~\ref{check}, and an ensemble $P_{Ed}$
where all blocked configurations \emph{with a given energy density}
are taken with equal weight.
Formally, we can write
\begin{equation} \label{ed-ensemble}
P_{Ed}(\sigma) \propto e^{- \beta_c E(\sigma)},
\end{equation}
where $\sigma$ is a blocked configuration and $E(\sigma)$ its
energy density, and $\beta_c = 1/T_c$ is the inverse of the
configurational temperature~(\ref{c-temp}).

The inverse temperature $\beta_c$ that appears in~(\ref{ed-ensemble})
is itself a function of the energy density $E$, and can be computed
using an explicit expression for the configurational entropy $S_c$
(see~\cite{dsgl} and references therein) 
\begin{equation} \label{entropy}
S_c(E) = E \ln{(-2E)} + \frac{1-E}{2} \ln{(1-E)} - \frac{1+E}{2} \ln{(1+E)},
\end{equation}
from which we obtain
\begin{equation} \label{temp}
\beta_c(E) = \frac{\text{d} S_c}{\text{d} E} =
\ln{ \left ( \frac{-2E}{\sqrt{1-E^2}} \right ) }.
\end{equation}

To derive~(\ref{entropy}), consider a finite chain of $N$
spins with periodic boundary conditions.
The blocked configurations such that exactly $n$ bonds are
unsatisfied (i.e., are between spins of opposite sign) have
energy density $E(\nu)=-(N-2n)/N=-(1-2 \nu)$, where $\nu=n/N$,
and their number is
\begin{equation}
{\cal N}(N,n) = {N-n \choose n}.
\end{equation}
Indeed, this is the number of ways of inserting $n$ unsatisfied
bonds between the $N-n$ satisfied ones, in such a way that the
unsatisfied bonds are isolated.
It is now easy to see that
$\frac{1}{N} \ln({\cal N}(N,n)) \to S_c(E(\nu))$ as $N,n \to \infty$
and $n/N=\nu$ is kept constant.

The translation-ergodicity of the model and the translation-invariance
of the energy density imply that, with $P_0$-probability $1$, in the
thermodynamic limit, the energy density of the final blocked
configuration is a deterministic constant, which can be computed by
solving exactly a dynamical equation for the densities of clusters of
consecutive unsatisfied bonds.
This is done in~\cite{dsgl} (see also~\cite{dl,ld,pb}), where the energy
density for the deep-quench situation that we are considering in this paper
(see~(\ref{initial})) is shown to be $E = -1 + e^{-1} \approx -0.63212$.
Therefore, using~(\ref{temp}), the appropriate value of the inverse
temperature in~(\ref{ed-ensemble}) for the deep-quench situation is
$\beta_c \approx 0.4895$.

Having an exact calculation for the configurational inverse
temperature $\beta_c$, we can use our general criterion to
check whether $P_{\infty} = P_{Ed}$ or not.
A simple calculation is sufficient to rule out this possibility.
Let
$\hat P_{Ed}(\cdot) = P_{Ed}(\cdot \, | \, \sigma_{\pm 2} = \sigma_{\pm 3} = +1)$
be the distribution $P_{Ed}$ conditioned on having $\sigma_{\pm 2} = \sigma_{\pm 3} = +1$.
Then, we have the following straightforward relations (with reference
to the events $A,B,C$, and $D$ of Section~\ref{check})
\begin{eqnarray}
\hat P_{Ed}(B) = e^{-2 \beta_c} \hat P_{Ed}(A), \\
\hat P_{Ed}(B) = \hat P_{Ed}(C) = \hat P_{Ed}(D), \\
\hat P_{Ed}(A) + \hat P_{Ed}(B) + \hat P_{Ed}(C) + \hat P_{Ed}(D) = 1,
\end{eqnarray}
from which it follows that
\begin{equation}
\hat P_{Ed}(A) = \frac{1}{1 + 3 e^{- 2 \beta_c}}.
\end{equation}
Identifying $P_{\infty}$ with $P_{Ed}$ would imply, using~(\ref{fkg}),
\begin{equation}
1 + 3 e^{- 2 \beta_c} \leq 2
\end{equation}
and thus
\begin{equation}
\beta_c \geq \frac{1}{2} \ln{3} \approx 0.5493,
\end{equation}
which contradicts the value $\beta_c \approx 0.4895$ corresponding
to the deep-quench situation that we are considering.

\bigskip
\bigskip

\noindent {\bf Acknowledgements.} The author thanks Frank den Hollander
and Marco Isopi for useful comments.


\bigskip

\begin{thebibliography}{99}

\bibitem{bray} A.~J.~Bray,
\emph{Adv.~Phys.} {\bf 43}, 357 
(1994).

\bibitem{gc} G.~Gallavotti, E.~G.~D.~Cohen,
\emph{Phys.~Rev.~Lett.} {\bf 74}, 2694 
(1995).

\bibitem{ckp} L.~F.~Cugliandolo, J.~Kurchan, L.~Peliti,
\emph{Phys.~Rev.~E} {\bf 55}, 3898 
(1997).

\bibitem{bckm} J.~P.~Bouchaud, L.~Cugliandolo, J.~Kurchan, M.~Mezard,
in \emph{Spin-glases and random fields} (ed. A.~P.~Young),
World Scientific, Singapore (1998).

\bibitem{fin} L.~R.~Fontes, M.~Isopi, C.~M.~Newman,
\emph {Probab.~Theory Relat.~Fields} {\bf 115}, 417 
(1999).

\bibitem{ruelle} D.~Ruelle,
\emph{Nature} {\bf 414}, 263 
(2001).

\bibitem{mk} H.~A.~Makse, J.~Kurchan,
\emph{Nature} {\bf 415}, 614 
(2002).

\bibitem{cdsn} F.~Camia, E.~De~Santis, C.~M.~Newman,
\emph{Ann.~Appl.~Probab.} {\bf 12}, 565 
(2002).


\bibitem{babg2} G.~Ben~Arous, A.~Bovier, V.~Gayrard,
\emph{Comm.~Math.~Phys.} {\bf 236}, 1 
(2003).

\bibitem{ed1} S.~F.~Edwards, in \emph{Granular Matter: An Interdisciplinary Approach}
(ed. A.~Mehta), 
Springer, New York (1994).

\bibitem{ed2} S.~F.~Edwards, in \emph{Disorder in Condensed Matter Physics}
(eds. J.~Blackman, J.~Taguena), 
Oxford~Univ.~Press, Oxford (1991).

\bibitem{me} A.~Mehta, S.~F.~Edwards, 
\emph{Physica~A} {\bf157}, 1091 
(1989).

\bibitem{bfs} J.~Berg, S.~Franz, M.~Sellitto,
\emph{Eur.~Phys.~J.~B} {\bf 26}, 349 
(2002).

\bibitem{fv} S.~Franz, M.~A.~Virasoro,
\emph{J.~Phys.~A} {\bf 33}, 891 
(2000).

\bibitem{dsgl} G.~De~Smedt, C.~Godr\`eche, and J.~M.~Luck,
\emph{Eur.~Phys.~J.~B} {\bf 27}, 363 
(2002).

\bibitem{harris1} T.~E.~Harris,
\emph{Proceedings~of~the~Cambridge~Phil.~Soc.} {\bf 56}, 13 
(1960).

\bibitem{fkg} C.~M.~Fortuin, P.~W.~Kasteleyn, J.~Ginibre,
\emph{Comm.~Math.~Phys.} {\bf 22}, 89 
(1971).

\bibitem{grimmett} G.~R.~Grimmett, \emph{Percolation},
Second edition, Springer, Berlin (1999).

\bibitem{dl} D.~S.~Dean, A.~Lef\'evre,
\emph{Phys.~Rev.~Lett.} {\bf 86}, 5639 (2001).

\bibitem{ld} A.~Lef\'evre, D.~S.~Dean,
\emph{J.~Phys.~A} {\bf 34}, L213 (2001).

\bibitem{pb} A.~Prados, J.~J.~Brey,
\emph{J.~Phys.~A} {\bf 34}, L453 (2001).

\bibitem{harris2} T.~E.~Harris,
\emph{Ann.~Probab.} {\bf 5}, 451 
(1975).

\bibitem{liggett} T.~M.~Liggett, \emph{Interacting Particle Systems}, Springer,
New York (1985).

\end{thebibliography}
\end{document}